\documentclass[pra,aps,amsmath,amssymb]{revtex4}
\usepackage{graphicx, subfigure}
\usepackage{subfigure}
\usepackage{color}

\newtheorem{thm}{Theorem}

\begin{document}
\title{Dissipative and Non-dissipative Single-Qubit Channels: Dynamics
  and Geometry}  \author{S. Omkar} \affiliation{Poornaprajna Institute
  of  Scientific Research,  Sadashivnagar, Bengaluru-  560080, India.}
\author{R. Srikanth} \affiliation{Poornaprajna Institute of Scientific
  Research,     Sadashivnagar,      Bengaluru-     560080,     India.}
\affiliation{Raman   Research  Institute,   Sadashivnagar,  Bengaluru-
  560060,  India}  \author{Subhashish  Banerjee} \affiliation{  Indian
  Institute of Technology Rajasthan, Jodhpur- 342011, India}

\begin{abstract}
Single-qubit channels  are studied under two  broad classes: amplitude
damping channels  and generalized depolarizing  channels.  A canonical
derivation of  the Kraus  representation of the  former, via  the Choi
isomorphism  is   presented  for  the  general  case   of  a  system's
interaction with  a squeezed thermal  bath.  This isomorphism  is also
used to characterize the difference  in the geometry and rank of these
channel classes.  Under the isomorphism, the degree  of decoherence is
quantified  according to  the mixedness  or separability  of  the Choi
matrix.  Whereas the  latter  channels form  a  3-simplex, the  former
channels  do  not  form  a  convex  set as  seen  from  an  ab  initio
perspective.   Further,  where the  rank  of generalized  depolarizing
channels can be any positive integer upto 4, that of amplitude damping
ones is either 2 or 4. Various channel performance parameters are used
to bring out the different  influences of temperature and squeezing in
dissipative channels. In particular, a noise range is identified where
the  distinguishability  of  states  improves  inspite  of  increasing
decoherence due to environmental squeezing.

\end{abstract} 

\pacs{03.67.-a,03.67.Hk,03.65.Yz} 

\maketitle

\section{Introduction}
Open quantum systems  are ubiquitous in the sense  that any system can
be thought  of as  being surrounded by  its environment  (reservoir or
bath) which influences its dynamics.  They provide a natural route for
discussing damping and dephasing. One of the first testing grounds for
open system  ideas was in quantum optics  \cite{wl73}. Its application
to other areas gained momentum  from the works of Caldeira and Leggett
\cite{cl83}, and Zurek \cite{wz93},  among others.

If the system and environment start out in product state, then
the  evolution of the  state $\rho$  can be  described by  the quantum
process $\rho ^\prime=\mathcal{E}(\rho)$. It  can be given an operator
sum representation or Kraus representation \cite{kraus,nc00,sudarshan}:
\begin{equation}
\mathcal{E}(\rho)=\sum_j E_j\rho E_j^{\dagger},
\end{equation}
where  $\sum_j  E_j^{\dagger}E_j=\cal{I}$.   The operators  $E_j$  are
called  Kraus   operators  or  the  operator   elements  of  operation
$\mathcal{E}$.  It  may be  noted that the  converse problem,  that of
deducing  the underlying  Lindbladian process  that generates  a given
completely   positive   (CP)  map   on   the   density  operator,   is
computationally  hard.  Complexity  theoretically, it  is known  to be
NP-hard \cite{cubitt}.

A result now familiar in quantum information theory is the isomorphism
between  the trace-preserving,  CP  maps on  a $d$-dimensional  system
(qudit)  and  the $d^4-d^2$  dimensional  space  of two-qudit  density
operators $\rho$  which are  maximally mixed on  one of  the particles
\cite{horodecki,jamio}.  One way to  obtain the state from the channel
is to  apply the latter on  one half of a  maximal two-qudit entangled
state.   The  resulting state  is  called  the  Choi matrix.   In  the
converse direction, a unique qudit channel can be associated with each
such Choi  matrix via the  notion of gate  teleportation \cite{vidal}.
It can be shown that the  Kraus operators for the qudit channel can be
derived by diagonalizing  the Choi matrix \cite{choi,debbie}, obtained
also  by  constructing  the   dynamical  map  for  the  transformation
\cite{rodrig,usha}.

In  this  work,  we  derive  the  Kraus  operators  for  the  squeezed
generalized amplitude damping (SGAD) channel in its canonical form via
the Choi  matrix method,  and establish its  unitary equivalence  to a
previous  derivation  \cite{srisub},   where  the  connection  to  the
amplitude  damping  (AD)   and  generalized  amplitude  damping  (GAD)
channels was manifest. The  channel-state isomorphism is used to study
and contrast  amplitude damping channels  and generalised depolarising
channels geometrically. Understanding the geometry of 1-qubit channels
is important as  it can simplify the study of  other problems, such as
channel  capacity  \cite{uhl1}.   Finally  the  contrastive  roles  of
temperature and squeezing in the case of the SGAD channel are noted.

\section{Some physically motivated single-qubit channels}

Depending upon the  system-reservoir ($S-R$) interaction, open systems
can  be   broadly  classified  into  two   categories,  viz.,  quantum
non-demolition (QND) or dissipative.  A particular type of (QND) $S-R$
interaction may be  achieved when the Hamiltonian $H_S$  of the system
commutes with the Hamiltonian $H_{SR}$ describing the system-reservoir
interaction, i.e., $H_{SR}$  is a constant of the  motion generated by
$H_S$ \cite{sgc96, mp98, gkd01,  subgho, bvt80}.  This results in pure
dephasing  without  dissipation.   Investigation into  pure  dephasing
scenarios was originally motivated by  the problem of the detection of
gravitational  waves  \cite{ct80, bo96}.   A  dissipative open  system
would be when $H_S$ and $H_{SR}$ do not commute resulting in dephasing
along with damping \cite{bp02}.   Impressive progress has been made on
the experimental front in the manipulation of quantum states of matter
towards  quantum  information  processing and  quantum  communication.
Myatt  {\it   et  al.}   \cite{myatt}  and  Turchette   {\it  et  al.}
\cite{turch}  performed   a  series  of  experiments   in  which  they
engineered  both the  pure dephasing  as well  as dissipative  type of
evolutions.   In  another  experiment  \cite{jb03}, a  QND  scheme  of
measurement was  characterized using  only linear optics  devices.  An
experimental  investigation  of the  dynamics  of  different kinds  of
bipartite  correlations,   in  an   all-optical  setup  was   made  in
\cite{xg10}.  In \cite{kp10}, an interesting experiment was presented,
in which dissipation induces  entanglement between two atomic objects.
Here we briefly discuss the  two processes, QND as well as dissipative
as applicable  to single  qubit channels.  In  this work we  model the
reservoir by a squeezed thermal  bath.  An advantage is that the decay
rate of quantum coherence can be suppressed leading to preservation of
nonclassical effects \cite{subgho,  kw88, kb93}.  A squeezed reservoir
may be  constructed on  the basis of  establishment of  squeezed light
field \cite{buz91}. Experiments probing the squeezed-light-atom system
have been  carried out in  Refs.  \cite{geo95,tur98}. All  the results
pertaining to  the usual thermal bath  can be obtained  by setting the
bath squeezing parameters to zero.

\subsection{QND Channel}

Following  \cite{subgho}, the  evolution equation  for a  system,  e.g. a
qubit,  interacting  with  its   environment  by  a  coupling  of  the
energy-preserving QND type where the  environment is a bosonic bath of
harmonic oscillators initially in  a squeezed thermal state, initially
decoupled from  the system,  in the system  eigenbasis denoted  by the
subscripts $n$, $m$, is:
 \begin{equation}
{d \over dt}\rho^s_{nm} (t) = \left[ -{i \over \hbar} (\epsilon_n - \epsilon_m) + i 
\dot{\eta} (t) (\epsilon^2_n - \epsilon^2_m) - (\epsilon_n - \epsilon_m)^2 \dot{\gamma} (t) 
\right] \rho^s_{nm} (t), \label{h2} 
\end{equation}
where
\begin{equation}
\eta (t) = - \sum\limits_k {g^2_k \over \hbar^2 \omega^2_k} 
\sin (\omega_k t), \label{h3} 
\end{equation}
and
\begin{equation}
\gamma (t) = {1 \over 2} \sum\limits_k {g^2_k \over \hbar^2 
\omega^2_k} \coth \left( {\beta \hbar \omega_k \over 2} \right) 
\left| (e^{i\omega_k t} - 1) \cosh (r_k) + (e^{-i\omega_k t} - 
1) \sinh (r_k) e^{2i \Phi_k} \right|^2. \label{h4} 
\end{equation}

Here  $\omega_k$ is  the  reservoir oscillator  frequency, indexed  by
subscript $k$, $\beta  = 1/{k_B T}$ and $g_k$  is the system-reservoir
coupling term.   For the case of zero  squeezing, $r = \Phi  = 0$, and
$\gamma  (t)$  given by  Eq.   (\ref{h4})  reduces  to the  expression
obtained  earlier \cite{sgc96,mp98,gkd01}  for the  case of  a thermal
bath. It can be seen that  $\eta (t)$ (\ref{h3}) is independent of the
bath initial conditions and hence  remains the same as for the thermal
bath.   Note that  in Eq.   (\ref{h2}), the  term responsible  for the
decay of  coherences, i.e., the  coefficient of $\dot{\gamma}  (t)$ is
dependent on  the eigenvalues  $\epsilon_n$ of the  `conserved pointer
observable'  operator which  in this  case is  the  system Hamiltonian
itself. This reiterates the observation that the decay of coherence in
a system  interacting with its bath  via a QND  interaction depends on
the  conserved pointer  observable  and the  bath coupling  parameters
\cite{mp98}. The  channel corresponding to the  evolution generated by
Eq. (\ref{h2}) is called  the phase flip channel \cite{subgho, srigp}.
More  generally,  phase  flip  channels  are  a  subset  of  Pauli  or
generalized depolarising channels, which are unital, i.e, they map the
identity matrix to itself.

\subsection{Dissipative Channel}

Consider  a  two-level  system  (qubit) interacting  with  a  squeezed
thermal  bath in  the weak  Born-Markov, rotating  wave approximation.
The system  Hamiltonian is  given by $H_S  = (\hbar\omega/2)\sigma_z$.
The system interacts with the reservoir via the atomic dipole operator
which in  the interaction  picture is given  as $\vec{D}(t)  = \vec{d}
\sigma_-  e^{-i\omega t}  +  \vec{d^*} \sigma_+  e^{i\omega t}$  where
$\vec{d}$ is  the transition matrix  elements of the  dipole operator.
The  master equation depicting  the evolution  of the  reduced density
matrix operator of  the system $S$ in the  interaction picture has the
following form \cite{bp02,srisub}
\begin{eqnarray}
{d \over dt}\rho^s(t) &=& \gamma_0 (N + 1) \left(\sigma_-  \rho^s(t)
\sigma_+ - {1 \over 2}\sigma_+ \sigma_- \rho^s(t) -
{1 \over 2} \rho^s(t) \sigma_+ \sigma_- \right) \nonumber\\
& + & \gamma_0 N \left( \sigma_+  \rho^s(t)
\sigma_- - {1 \over 2}\sigma_- \sigma_+ \rho^s(t) -
{1 \over 2} \rho^s(t) \sigma_- \sigma_+ \right) \nonumber\\
& - & \gamma_0 M   \sigma_+  \rho^s(t) \sigma_+ -
\gamma_0 M^* \sigma_-  \rho^s(t) \sigma_-. 
\label{4b} 
\end{eqnarray}
Here $\gamma_0$ is the spontaneous  emission rate given by $\gamma_0 =
(4  \omega^3 |\vec{d}|^2)/(3 \hbar  c^3)$, and  $\sigma_+$, $\sigma_-$
are the standard raising and lowering operators, respectively given by
$\sigma_+ =  |1 \rangle  \langle 0| =  \frac{1}{2} \left(\sigma_x  + i
\sigma_y \right)$ and $\sigma_- = |0  \rangle \langle 1| = {1 \over 2}
\left(\sigma_x - i \sigma_y \right)$.  Eq. (\ref{4b}) may be expressed
in a manifestly Lindblad form as \cite{srigp}
\begin{equation}
\frac{d}{dt}\rho^s(t) = \sum_{j=1}^2\left(
2R_j\rho^s R^{\dag}_j - R_j^{\dag}R_j\rho^s - \rho^s R_j^{\dag}R_j\right),
\label{eq:lindblad}
\end{equation}
where    $R_1   =    (\gamma_0(N_{\rm   th}+1)/2)^{1/2}R$,    $R_2   =
(\gamma_0  N_{\rm  th}/2)^{1/2}R^{\dag}$  and  $R =  \sigma_-\cosh(r)  +
e^{i\Phi}\sigma_+\sinh(r)$.  This guarantees that the evolution of the
density operator  is CP. If $T=0$,  then $R_2$ vanishes,  and a single
Lindblad operator suffices to describe Eq. (\ref{4b}). Also
\begin{equation}
\label{eq:N}
N   =  N_{\rm   th}(\cosh^2(r)  +   \sinh^2(r))  +
\sinh^2(r),
\end{equation}
and $M =  -\frac{1}{2} \sinh(2r) e^{i\Phi} (2 N_{\rm  th} + 1)$.  Here
$N_{\rm  th}  =  1/(e^{\hbar  \omega/k_B   T}  -  1)$  is  the  Planck
distribution  giving the number  of thermal  photons at  the frequency
$\omega$; $r$  and $\Phi$ are bath squeezing  parameters.  The general
map   generated  by   the  Eq.    (\ref{4b})  is   the   SGAD  channel
\cite{srisub}, which generalizes the notion of the AD and GAD channels
\cite{nc00}.   These  amplitude  damping  channels are  non-unital  and
contractive, mapping any initial state to a unique asymptotic state.

\section{Some properties of the Kraus representation of
dissipative and non-dissipative channels}

A superoperator  ${\cal E}$ due  to interaction with  the environment,
acting on the state of the system is given by
\begin{equation}
\label{eq:kraus}
\rho  \longrightarrow  {\cal  E}(\rho)  =  \sum_k  \langle  e_k|U(\rho
\otimes |0\rangle\langle 0|)U^{\dag}|e_k\rangle  = \sum_j E_j \rho
E_j^{\dag},
\end{equation}
where $U$ is  the unitary operator representing the  free evolution of
the system,  reservoir, as  well as the  interaction between  the two,
$|0\rangle$ is the  environment's initial state, and $\{|e_k\rangle\}$
is a  basis for  the environment. The  environment and the  system are
assumed  to  start  in  a  product  state.  The  $E_j  \equiv  \langle
e_k|U|0\rangle$   are   the  Kraus   operators,   which  satisfy   the
completeness condition $\sum_j E_j^{\dag}E_j = \mathcal{I}$. It can be
shown  that  any   transformation  that  can  be  cast   in  the  form
(\ref{eq:kraus}) is a CP map \cite{nc00}.

There are  infinitely many Kraus operator  representations even within
the same representation  basis of the system, depending  on the choice
of tracing basis $\{|e_k\rangle\}$  of the environment.  Each of these
sets  of  Kraus operators  is  unitarily  related  to the  other:  let
$E_k=\langle       e_k|U|0\rangle$       and       $E^\prime_k=\langle
e^\prime_k|U|0\rangle$.   Define  unitary   operation  $V$  such  that
$\langle  e^\prime_k|=\langle e_k|V^{\dagger}$, and  hence $E^\prime_k
=\langle                e_k|V^{\dagger}U|0\rangle$.                Now
$V|e_k\rangle=\sum_{j}\alpha_{j,k}|e_j\rangle$ and thus
\begin{equation}
 E^\prime_k=\langle e_k|V^{\dagger}U|0\rangle=
 \sum_{j}\alpha^{\ast}_{j}\langle e_j|U|0\rangle=
 \sum_{j}\alpha^{\ast}_{j,k}E_j.
\end{equation}
The above can be represented as a matrix-valued vector equation
\begin{equation}
\vec{E^\prime_k}=V^{\dagger}\vec{E_k}.
\label{eq:unit}
\end{equation}

Let $\vec{A}=(\hat{A}_i)$  and $\vec{B}=(\hat{B}_i)$ where $\hat{A}_i$
and $\hat{B}_i$  are $d\times d$  matrices (here Kraus  operators) and
$i=1,2,\dots,d^2$.  Consider the  transformation of  the matrix-valued
inner product between $\vec{A}$ and $\vec{B}$.
\begin{eqnarray}
\vec{A}^\prime.\vec{B}^\prime 
   &=& \sum_i (\hat{A}_i^{\prime})^{\dagger}\hat{B}_i^\prime\nonumber\\
&=& \sum_{ijk} (U_{ij}\hat{A}_j)^\dagger U_{ik}\hat{B}_k\nonumber\\
&=& \sum_{ijk} \hat{A}_j^\dagger U_{ji}^* U_{ik}\hat{B}_k\nonumber\\
&=& \sum_{jk} \hat{A}_j^\dagger \delta_{jk}\hat{B}_k\nonumber\\
&=& \sum_{j}\hat{A}_j^\dagger\hat{B}_j.\nonumber\\
\end{eqnarray}
As a corollary, the Hilbert-Schmidt product of any two Kraus `vectors', 
$\sum_j {\rm Tr}(A^\dag_j B_j)$ is preserved. 

Consider  the vector  obtained by  reading  off a  fixed element,  say
element of index $lm$,  namely, $(E_j)_{lm}$, for each Kraus operator.
Eq.   (\ref{eq:unit})  can  be   thought  of  as  applying  a  unitary
transformation  to $d^2$ (not  necessarily independent)  such vectors.
Thus one  can define a  host of other  norms that are  preserved under
this transformation.   Any channel  parameter (such as  gate fidelity,
etc.) must  be a function  of such a  generalized norm in order  to be
unitarily invariant  under the transformation  Eq. (\ref{eq:unit}) and
thus  be a  valid measure  to characterize  channel  performance.  For
example,  the   quantity  $\sum_j  |\textrm{Tr}(E_j)|^2$   is  another
acceptable norm.

The  Kraus operators  $K_j$ obtained  by the  Choi method  satisfy the
orthogonality condition Tr$(E^\dag_jE_k)=0$ for $j\ne k$, which is not
a unitarily  invariant condition.  In particular,  the Kraus operators
for the SGAD channel obtained in Ref. \cite{srisub} lack this form.

The  SGAD channel  derived  here is  typical  of dissipative  channels
\cite{subgho}, which are characterized by the non-commutativity of the
interaction Hamiltonian $H_{SR}$ and  the system Hamiltonian $H_S$. By
contrast, the QND case, where these  two do commute, is marked by pure
phase damping and no  dissipation, i.e., populations remain unchanged.
Here  we   show  that  the  condition   $[H_S,H_{SR}]=0$  implies  the
commutativity  of the  Kraus  operators and  quantum  states used  for
communication (the signal states), assumed to be eigenstates of $H_S$.
Let $|e\rangle$ be the initial  state of the environment (which may be
generalized  to  a separable  mixed  state)  and $\{|e_j\rangle\}$  an
environmental basis.

For  arbitrary   non-dissipative  interaction,  we   take:  $H_{SR}  =
\sum_k\alpha_k|k\rangle\langle         k|\otimes\hat{P}$,        where
$\{|k\rangle\}$ is  a basis  for the first  particle and  $\hat{P}$ an
environmental observable. Given $H = H_S  + H_R + H_{SR}$, with $H_S =
\sum_k\lambda_k |k\rangle\langle k|$ and $H_R = f(\hat{P})$, we have:
\begin{eqnarray}
E_j&=&\langle e_j|e^{iH t}|e\rangle\nonumber\\
&=&\langle e_j|e^{i\sum_k(\lambda_k|k\rangle\langle k| + \alpha_k|k\rangle\langle k|\otimes\hat{P} t)}|e^\prime\rangle\nonumber\\
&=&\langle e_j|\sum_k| k\rangle\langle k|\otimes  
e^{i(\lambda_k + \alpha_k\hat{P} t)}|e^\prime\rangle\nonumber\\
&=&\sum_k| k\rangle\langle k|\beta_k^{(j)},
\end{eqnarray}
where   $\beta^{(j)}_k   \equiv    \langle   e_j|   e^{i(\lambda_k   +
  \alpha_k\hat{P}   t)}|e^\prime\rangle$   and   $|e^\prime\rangle   =
e^{if(\hat{P})}|e\rangle$.    If   the   eigenstates   of   the   system
  Hamiltonian,  denoted $\{|k\rangle\}$,  are taken  to be  the signal
  states, then the statement that  $[|k\rangle\langle k|, E_j] = 0$ is
  equivalent to $[H_S,H_{SR}]=0$.  If the signal states do not commute
  with   Kraus    operators,   then   $[H_S,H_{SR}]\ne0$,    and   the
  system-environmental  interaction must  be dissipative.   This  is a
  unitarily invariant feature  since the condition $[H_S,H_{SR}]=0$ is
  independent  of  the  tracing  basis  used to  determine  the  Kraus
  operators.

\section{Canonical Kraus representation of the SGAD channel \label{sec:sgad}}

The  action of  the SGAD  channel on  the single  qubit  state $\rho$,
denoted $\mathcal{E}$, is given by\cite{srisub}:
\begin{eqnarray}
\langle      \sigma_x(t)\rangle      &=&      [1+\frac{1}{2}(e^{\gamma_0
    at}-1)(1+\cos(\Phi))]e^{\frac{-\gamma_0(2N+1+a)t}{2}}\langle
\sigma_x(0)\rangle-                        \sin(\Phi)\sinh(\frac{\gamma_0
  at}{2})e^{\frac{-\gamma_0                           (2N+1)t}{2}}\langle
\sigma_y(0)\rangle\nonumber         \\        &\equiv&        A\langle
\sigma_x(0)\rangle-B\langle  \sigma_y(0)\rangle, \nonumber  \\ \langle
\sigma_y(t)\rangle             &=&            [1+\frac{1}{2}(e^{\gamma_0
    at}-1)(1-\cos(\Phi))]e^{\frac{-\gamma_0         (2N+1+a)t}{2}}\langle
\sigma_y(0)\rangle-                        \sin(\Phi)\sinh(\frac{\gamma_0
  at}{2})e^{\frac{-\gamma_0                          (2N+1)t}{2}}\langle
\sigma_x(0)\rangle\nonumber         \\        &\equiv&        G\langle
\sigma_y(0)\rangle-B\langle  \sigma_x(0)\rangle,\nonumber  \\  \langle
\sigma_z(t)\rangle     &=&    e^{-\frac{-\gamma_0    (2N+1)t}{2}}\langle
\sigma_z(0)\rangle-                          \frac{(1-e^{-\frac{-\gamma_0
      (2N+1)t}{2}})}{2N+1}\equiv H\langle \sigma_z(0)\rangle-Y.
\label{eq:sgad}
\end{eqnarray}
Here $N$, $\gamma_0$, $r$, $\Phi$ are as defined in Eq. (\ref{eq:N}) and $a = \sinh(2r)(2 N_{th} + 1)$.

Consider the maximally entangled (unnormalized) state $|\tilde{\psi}\rangle=(|00\rangle+ |11\rangle)$.
We find, using Eq. (\ref{eq:sgad}), the Choi matrix
\begin{equation}
C_{\mathcal{E}} \equiv (I\otimes\mathcal{E})
|\tilde{\psi}\rangle\langle\tilde{\psi}| = 
\left(\begin{array}{clclr}
\left(\begin{array}{clclr}(1+ H - Y)/2&0\\0&(1-H + Y)/2\\ \end{array}\right)&
\left(\begin{array}{clclr}0&\frac{A+G}{2}\\(\frac{A-G}{2}-iB)&0\\ \end{array}\right)\\
\left(\begin{array}{clclr}0&(\frac{A-G}{2}+iB)\\\frac{A+ G}{2}&0\\ \end{array}\right)&
\left(\begin{array}{clclr}(1-H-Y)/2&0\\0&(1+ H + Y)/2\\ \end{array}\right)\\
\end{array}\right).
\label{eq:choi}
\end{equation}
According  to  Choi's  theorem,  the  $d^2$  Kraus  operators  can  be
constructed  by   `squaring'  (juxtaposing  $d$-element   long  column
segments  of)  the  eigenvectors  of $C_{\cal{E}}$,  which  have  been
normalized to  their eigenvalues \cite{debbie}.  They can be  shown to
be:
\begin{eqnarray}
J_\pm &=& \frac{1}{M_\pm}\left(\begin{array}{clclr}
0&\sqrt{1-H\mp \Psi}\\
i\frac{(\sqrt{1-H \mp \Psi})(\Psi \pm Y)}{2B+i(G-Aa)}&0\\
\end{array}\right), \nonumber\\
K_\pm &=& \frac{1}{N_\pm}\left(\begin{array}{clclr}
\frac{-\sqrt{1+H \mp \eta}(Y \pm \eta)}{A+G}&0\\
0&\sqrt{1+H \mp \eta}\\
\end{array}\right), \nonumber\\
\label{krauschoi}
\end{eqnarray}
where $\Psi=\sqrt{(A-G)^2+4B^2+Y^2},$ 
$\eta=\sqrt{(A+G)^2+Y^2},$ 
and
$ M_\pm=\sqrt{2}\sqrt{1+\left|\frac{\mp Y+\Psi}{2B+i(G-A)}\right|^2},$\\
$ N_\pm=\sqrt{2}\sqrt{1+ \left|\frac{Y\pm\eta}{A+G}\right|^2}$.

The  Kraus operators  for the  noise  process, generated  by the  SGAD
channel, were derived in Ref.  \cite{srisub}, using an ansatz based on
a standard operator-sum representation \cite{nc00}.  As illustrated by
the Eq. (\ref{eq:unit}), a  necessary and sufficient condition for the
equivalence of  two different  Kraus operator representations  is that
they are related by a  unitary transformation. We demonstrate this for
the SGAD channel by  finding the unitary transformation connecting the
Kraus  operators derived via  Choi formalism,  Eqs. (\ref{krauschoi}),
and   those   derived  in   Ref.   \cite{srisub},   which  we   denote
$J_{\pm}^\prime$ and $K_{\pm}^\prime$. Writing
\begin{equation}
\left( \begin{array}{c} J_+^\prime \\ J_-^\prime \\ K_+^\prime \\ 
K_-^\prime \end{array} \right)
= U
\left( \begin{array}{c} J_+ \\ J_- \\ K_+ \\ K_- \end{array}  \right), \label{unitequiv}
\end{equation}
we find that
\begin{equation}
U = \left(\begin{array}{clclr}
0& \Upsilon_+ & 0& 
\Upsilon^{\prime\prime}_+ \\
0 & \Upsilon_- &0&
\Upsilon^{\prime\prime}_- \\
\Upsilon^\prime_+ &0&
\tilde{\Upsilon}_+ &0\\
\Upsilon^\prime_- &0&
\tilde{\Upsilon}_- &0\\
\end{array}\right),
\end{equation}
where
\begin{eqnarray}
\Upsilon_\pm &=& \frac{\sqrt{1-H\mp\Psi}}{M_\pm\sqrt{p_1\alpha}}
\left(\frac{\pm i(\Psi \mp Y)}{2B+i(G-A)}-\sqrt{\frac{\mu}
{\nu}}e^{-i\theta}\right), 
\nonumber \\
\Upsilon^\prime_\pm &=& \frac{\sqrt{1+H \mp \eta}}{N_\pm\sqrt{p_1}(\sqrt{1-\mu}-\sqrt{1-\alpha}\sqrt{1-\nu})}\left(\sqrt{1-\mu}-
\frac{\sqrt{1-\nu}(Y\pm\eta)}{A+G}\right) ,\nonumber \\
\Upsilon^{\prime\prime}_\pm &=& \frac{\sqrt{1-H\mp \Psi}}{M_\pm\sqrt{p_2\nu}}, \nonumber \\
\tilde{\Upsilon}_\pm &=& -\frac{\sqrt{1+H \mp \eta}}{N_\pm\sqrt{p_2}
\left(\sqrt{1-\mu}-\sqrt{1-\alpha}\sqrt{1-\nu}\right)}\left(\sqrt{1-\alpha}-\
\frac{(Y\pm \eta)}{A+G}\right). 
\end{eqnarray}

Also, the terms $\mu$, $\nu$,  $\theta$, $\alpha$, $p_1$ and $p_2$ are
as defined  in Eqs. (28) to  (32) of \cite{srisub}. It  may be checked
that $U U^{\dagger}= U^{\dagger}U = \mathcal{I}$.

\section{Geometric structure of channels}

 Given  any  set of  points  $x_i\in  S$,  if the  convex  combination
 $x=\sum_i \mu_i x_i \in S$,  where $\mu_i \ge 0$ and $\sum_i\mu_i=1$,
 then the set $S$  is {\it convex}. A point $x$ is  said to be pure or
 extreme  if  it  cannot   be  expressed  as  a  (non-trivial)  convex
 combination two  or more  points.  The smallest  convex set  $H$ that
 contains a given set $S$ is  the {\it convex hull} of $S$. The convex
 hull  of a  given  finite number  of  pure points  is  a {\it  convex
   polytope}.   Geometrically,  a polytope  can  be  visualized as  an
 object or tile  with flat sides.  In the space  of dimension $n$, the
 convex  hull  of  $n+1$ points  that  are  not  confined to  a  $n-1$
 dimensional subspace is an  $n$-simplex, $\Xi_n$. The {\it dimension}
 of  a given  convex set  $S$ is  the largest  integer $n$,  such that
 $\Xi_n\in S$.

\subsection{Channel rank}

Given  a map  $\Phi$ that  maps the  algebra of  $m \times  m$ complex
matrices  to another matrix  algebra, we  may define  the rank  of the
channel  as that  of the  matrix associated  with  $\Phi$ \cite{uhl2}.
Here,  by virtue of the Choi isomorphism,  one may associate  a rank
with  the channel,  identified  with that  of  the corresponding  Choi
matrix.  For the SGAD channel,  the eigenvalues of the Choi matrix are
given by
\begin{eqnarray}
e_\pm  &=&  \frac{1}{2}\left(1  - H  \pm  \sqrt{(A  -  G)^2 +  4B^2  +
  Y^2}\right),  \nonumber  \\  f_\pm  &=& \frac{1}{2}\left(1  +  H  \pm
\sqrt{(A + G)^2 + Y^2}\right).
\label{eq:rank}
\end{eqnarray}
Clearly,  $e_+  \ge  e_-$  and   $f_+  \ge  f_-$.   The  trivial  case
corresponds  to  the   unitary  channel,  wherein  channel  parameters
$T=r=0$,  and $f_+  = 1$  with all  other eigenvalues  equal to zero.   Let us
consider a nontrivial noise where $e_-  = 0$ for a given channel. From
Eq. (\ref{eq:rank}), it follows that $1 - H = \sqrt{(A - G)^2 + 4B^2 +
  Y^2}$.   This,  as  can   be  seen  from  Eq.   (\ref{eq:sgad}),  is
equivalent to $r=T=0$, which in turn  implies that $1 + H = \sqrt{(A +
  G)^2 + Y^2}$ and therefore that $f_-=0$. Conversely, it can be shown
that $f_- =  0 \Longrightarrow e_-=0$.  One way  to understand this is
to note  that Eq. (\ref{eq:lindblad}) that generates  the SGAD channel
${\cal E}$ has only one Lindblad  operator when $T=0$, and two when $T
> 0$.

We thus find  that $e_-$ and $f_-$ simultaneously  vanish (in the case
of unitary and amplitude damping  channels with vanishing $T$ and $r$)
or both  are non-vanishing (for  more general channels). For  the SGAD
channel for  a qubit, we  thus find  that the rank  is either 2  or 4.
This of  course is not a  general quantum feature,  and noise channels
for qubits exist with odd rank greater  than 1. An example of a rank 3
channel is the  Pauli channel with Kraus operator  elements $I, \sigma_x$ and
$\sigma_y$ with weights  $p, q$ and $r$, where $p+q+r=1$. 
For this the Choi matrix is given by:
\begin{equation}
\frac{1}{2}\left( \begin{array}{cccc}
p & 0 & 0 & p \\
0 & q+r & q-r & 0 \\
0 & q-r & q+r & 0 \\
p & 0 & 0 & p \end{array}\right),
\end{equation}
which  is  manifestly  of rank  3  (in that  precisely  3 rows  are  linearly
independent). 

\subsection{Pauli and depolarizing channels \label{sec:phf}}

Under the above  isomorphism, the set of unitaries on  a qudit maps to
pure states in $V$, the set  of two-qudit states isomorphic to CP maps
on a single qudit.  The  general state of a two-qubit density operator
is given by:
\begin{equation}
\rho  = \frac{1}{4}\left(I\otimes I + \sum_j r_j \sigma_j \otimes I_2
+ s_j I_2 \otimes \sigma_j + \sum_{j,k} t_{j,k} \sigma_j \otimes \sigma_k\right),
\end{equation}
where $r_j, s_j$ and the tensor $t_{j,k}$ are generally complex numbers subject to requirement
$\rho=\rho^{\dagger}$ and $Tr(\rho)=1$.
Letting
$|\psi\rangle = \frac{1}{\sqrt{2}}(|00\rangle + 
|11\rangle$, we have
\begin{equation}
|\psi\rangle\langle\psi| = \frac{1}{4}\left(I\otimes I+\sigma_x\otimes\sigma_x - \sigma_y\otimes\sigma_y + \sigma_z\otimes\sigma_z\right)
=
(\mathcal{E_I}\otimes I)|\psi\rangle\psi| 
\equiv \mathcal{I},
\end{equation}
where $\mathcal{E}_{\cal{I}}$  is the trivial  noise, corresponding to
the identity  operator. Under  the Choi isomorphism  the corresponding
state is therefore $\mathcal{I}$.

In this Section,  for two-qubit states which have  only the $I \otimes
I$  and  the $t_{j,j}$  components  non-vanishing,  we will  represent
$\rho$  by  its  \textit{signature},  the  list  of  these  components
multiplied  by 4. Thus,  $\mathcal{I} \equiv  \overline{(1, 1,-1,1)}$.
More generally,  the class  of states we  consider in  this subsection
have  the signature $\overline{(1,a,b,c)}$,  and are  characterized by
the   (quadratic)   mixedness   $\frac{d}{d-1}(1-\textrm{Tr}\rho^2)=1-
\frac{|a|^2+|b|^2+|c|^2}{3}$.

Consider  the phase  flip quantum  channel represented  by the  set of
Kraus operators [$\sqrt\alpha  I, \sqrt(1-\alpha)Z$], where $Z$ stands
for  the Pauli  operator $\sigma_z$  and $\alpha$  is a  real positive
number  such that $0\leq\alpha\leq  1$.  The  state isomorphic  to the
channel  corresponding to  application of  $Z$ is  $\mathcal{Z} \equiv
(\mathcal{E_Z}\otimes   I)|\tilde{\psi}\rangle\langle\tilde{\psi}|   =
\overline{(1,-1,1,1)}$.  Thus  the phase flip channel is  given by the
1-simplex, $\hat{\cal{F}}$:
\begin{equation}
\alpha\mathcal{I} + (1-\alpha)\mathcal{Z} =
\overline{(1,0,0,1)} +
(2\alpha-1)\overline{(0,1,-1,0)}.
\label{eq:alfa}
\end{equation}

It is closely  related to the phase damping channel,  given by the set
of       Kraus-operators:      $\left[\sqrt\beta      I,\sqrt(1-\beta)
  P_{0},\sqrt(1-\beta)  P_{1}\right]$,  where  $P_{0}=|0\rangle\langle
0|$  and $P_{1}=|1\rangle\langle  1|$  are projectors and $\beta$ is a
real positive number such that $0\leq\beta\leq 1$ .   By the Choi
isomorphism, they  correspond to  states: $\mathcal{I}$
and        $\mathcal{Z_P}        \equiv        (\mathcal{E_{P}}\otimes
I)|\tilde{\psi}\rangle\langle\tilde{\psi}|   =  \overline{(1,0,0,1)}$.
The phase  damping channel  is given by  the 1-simplex
\begin{equation}
\beta\mathcal{I} + (1-\beta)\mathcal{Z_P} =
\overline{(1,0,0,1)} + \beta\overline{(0,1,-1,0)}.
\label{eq:beta}
\end{equation}
Comparing this  with Eq.  (\ref{eq:alfa}), it is  seen that  the phase
damping channel  is strictly  a subset of  the phase flip  channel. In
particular,  the phase  damping  channel extreme  point obtained  with
$\beta=0$ corresponds  to the  mixed point of  the phase  flip channel
given by  $\alpha=\frac{1}{2}$.  The former corresponds  to the latter
in  the range  $\alpha \in  [\frac{1}{2},0]$, where  they  are related
according to $\beta =  2\alpha-1$.  

The generalised  depolarising or  Pauli channels have  Pauli operators
(apart   from   a   factor)    as   their   Kraus   operators,   i.e.,
$\left[\sqrt{\alpha}
  I,\sqrt{\beta}\sigma_{x},\sqrt{\gamma}\sigma_y,\sqrt{\delta}\sigma_z\right]$,
where  $\alpha,  \beta,  \gamma,   \delta  \ge  0$  are  real  numbers
satisfying   $\alpha+\beta+\gamma+\delta=1$.  We  define  $\mathcal{X}
\equiv                                            (\mathcal{E_X}\otimes
I)|\tilde{\psi}\rangle\langle\tilde{\psi}|   =  \overline{(1,1,1,-1)}$
and          $\mathcal{Y}         \equiv         (\mathcal{E_Y}\otimes
I)|\tilde{\psi}\rangle\langle\tilde{\psi}| = \overline{(1,-1,-1,-1)}$.
Thus every  Pauli channel is  a member of  the polytope given  by four
pure points $\mathcal{I}, \mathcal{X}, \mathcal{Y}, \mathcal{Z}$, as
\begin{equation}
\label{eq:abcd}
v = \alpha\mathcal{I} + \beta\mathcal{X} + \gamma\mathcal{Y}
+ \delta\mathcal{Z} =
\overline{(1,\alpha+\beta-\gamma-\delta,
-\alpha+\beta-\gamma+\delta,\alpha-\beta-\gamma+\delta)}.
\end{equation}
It   follows   from  the
properties  of  vector   spaces  that  if  $\mathcal{I},  \mathcal{X},
\mathcal{Y}$ and $\mathcal{Z}$  are mutually orthogonal, then the 
decomposition (\ref{eq:abcd}) is indeed unique.

It is  readily seen  that six inner  products between  these elements,
given   by   the    Hilbert-Schmidt   product   Tr($\mathcal{I   X}$),
Tr($\mathcal{Y X}$), etc.,  indeed vanish.  Thus the set  of all Pauli
channels,  the   polytope  $\mathcal{\hat{P}}$,  is   a  3-simplex  (a
tetrahedron)   embedded   within   $V$.    The  phase   flip   channel
$\mathcal{\hat{F}}$    corresponds    to    a   proper    subset    of
$\mathcal{\hat{P}}$,        in       particular,        the       edge
$(\mathcal{I},\mathcal{Z})$  of the  latter, and  the volume  of phase
damping  channels in this  set is  $\frac{1}{2}$.  This  structure has
been studied using  affine maps on Bloch sphere  in Ref. \cite{geetu},
where  it was  shown that  the fraction  of the  channels that  can be
simulated with a one-qubit environment is $\frac{3}{8}$.

The elements of the  important, depolarizing channel are characterized
by the action:
\begin{equation}
\rho \mapsto p\rho + (1-p)\frac{I}{2}.
\label{eq:depol}
\end{equation}
Noting that since for any $\rho$,  $\frac{I}{2} = \frac{1}{4}(\rho + \sigma_x\rho \sigma_x
+ \sigma_y\rho  \sigma_y + \sigma_z\rho \sigma_z)$, Eq.  (\ref{eq:depol}) is seen to  have a Kraus
representation             $\left[\sqrt{\frac{1+3p}{4}}             I,
  \sqrt{\frac{3(1-p)}{4}}\sigma_{x},   \sqrt{\frac{3(1-p)}{4}}\sigma_y,
  \sqrt{\frac{3(1-p)}{4}}\sigma_z\right]$.   The Choi matrix  for this
process has the convex structure:
\begin{eqnarray}
\mathcal{V}   &=&  p\mathcal{I}   +  \frac{(1-p)}{4}(   \mathcal{I}  +
\mathcal{X} + \mathcal{Y} + \mathcal{Z})  ~(0 \le p \le 1)  \nonumber
\\ &=& \overline{(1,p,-p,p)},
\label{eq:depolz}
\end{eqnarray}
which   is  just   the   two-qubit  Werner   state   $pI\otimes  I   +
(1-p)|\tilde{\psi}\rangle  \langle\tilde{\psi}|$.  

The twirling operation  \cite{horodecki} on states is defined by:
\begin{equation}
\mathcal{T}(\rho) \equiv \int dU U\otimes U^\ast \rho
U^\dag\otimes U^{\dag\ast}.
\end{equation}
While  it leaves  a  singlet  state invariant,  it  maps an  arbitrary
two-qubit  state to  a Werner  state. Interpreted  as an  operation on
maps, it maps any channel to the depolarizing channel. It can be shown
to have  the property that the  fidelity $F=F(|\tilde{\psi}\rangle, (I
\otimes   \mathcal{E})|\tilde{\psi}\rangle\langle|\tilde{\psi}|)$   is
preserved.  Under the  Choi isomorphism, this is a  contractive CP map
collapsing arbitrary  points in the above Pauli  3-simplex into points
in the depolarizing simplex.
 
For  the depolarizing  channels, representing  Werner family  of states
$\phi_D(p)    \equiv   \overline{(1,p-p,p)}$,    one   finds    $F   =
\frac{3p+1}{4}$, while for the Pauli channel, represented by the state
$\phi_P(\alpha,                  \beta,\gamma)                  \equiv
\overline{(1,\alpha+\beta-\gamma-\delta,
  -\alpha+\beta-\gamma+\delta,\alpha-\beta-\gamma+\delta)}$, one finds
$F  =  \alpha$, independent  of  $\beta,  \gamma,  \delta$.  From  the
property of preservation of $F$ under twirling, it follows $\phi_P$ is
twirled        to       $\phi_D        =       (1,\frac{4\alpha-1}{3},
-\frac{4\alpha-1}{3},\frac{4\alpha-1}{3})$.

It   follows   from   Eq.   (\ref{eq:depolz})  that   $\mathcal{V}   =
\alpha\mathcal{I} +  (1-\alpha)\mathcal{D}$ with, $(\frac{1}{4}  \le p
\le  1)$,  where  $\mathcal{D}  \equiv (\mathcal{X}  +  \mathcal{Y}  +
\mathcal{Z})/3$ and $\alpha  = (3p+1)/4$.  The set $\mathcal{\hat{D}}$
of  all  depolarizing  channels  forms  a  1-simplex  embedded  within
$\mathcal{\hat{P}}$,   suspended   from   $\mathcal{I}$  towards   the
$\mathcal{XYZ}$  base of  the tetrahedron,  but terminating  above the
base at the point $\frac{1}{4}(\mathcal{I} + \mathcal{X} + \mathcal{Y}
+ \mathcal{Z})$.

\subsection{The SGAD channel}

The  Choi matrix for  the generalized  amplitude damping  channel (the
SGAD  channel with squeezing  set to  zero) can  be obtained  from the
Kraus operators:
\begin{eqnarray}
\label{eq:gbmakraus}
\begin{array}{ll}
E_1 \equiv \sqrt{p}\left[\begin{array}{ll} 
\sqrt{1-\alpha} & 0 \\ 0 & 1
\end{array}\right]; ~~~~ &
E_2 \equiv \sqrt{p}\left[\begin{array}{ll} 0 & 0 \\ \sqrt{\alpha} & 0
\end{array}\right];  \\
E_3 \equiv \sqrt{1-p}\left[\begin{array}{ll} \sqrt{1-\mu} & 0 \\ 0 & 
\sqrt{1-\nu}
\end{array}\right]; ~~~~ &
E_4 \equiv \sqrt{1-p}\left[\begin{array}{ll} 0 & \sqrt{\nu} \\ \sqrt{\mu}e^{-i\phi} & 0
\end{array}\right],
\end{array}
\end{eqnarray}
where $0 \le p \le 1$ \cite{nc00,srisub}. 

The two corresponding Choi matrices, with Kraus operators $E_{1,2}$
and $E_{3,4}$, respectively, are:
\begin{eqnarray}
(\mathcal{E}_{12}\otimes I)|\tilde{\psi}\rangle\langle\tilde{\psi}| &=& \frac{1}{4}\left(I\otimes I-\sigma_z\otimes I+\sqrt{1-\alpha}\sigma_x\otimes\sigma_x -\sqrt{1-\alpha} \sigma_y\otimes\sigma_y + (1-\alpha)\sigma_z\otimes\sigma_z\right), \\ \nonumber
(\mathcal{E}_{34}\otimes I)|\tilde{\psi}\rangle\langle\tilde{\psi}| &=&
\frac{1}{4}\left(I\otimes I+(\nu-\mu)\sigma_z\otimes I+ (\sqrt{(1-\mu)(1-\nu}+ \sqrt{\mu\nu}\cos(\phi))\sigma_x\otimes\sigma_x
-\sqrt{\mu\nu}\sin(\phi) \sigma_y\otimes\sigma_x \right)\\ \nonumber
&&-\frac{1}{4}\left( \sqrt{1-\mu}I\otimes\sigma_y +\sqrt{\mu\nu}\cos(\phi)\sigma_y\otimes\sigma_y 
+\sqrt{\mu\nu}\sin(\phi)\sigma_x\otimes\sigma_y  \right).
\end{eqnarray}

It would seem from this that the generalized amplitude damping channel
(and by extension SGAD), has the convex structure
\begin{equation}
\Lambda = p{\mathcal {\hat{E}}_{12}} + (1-p)
{\mathcal {\hat{E}}_{34}},
\label{eq:naiv}
\end{equation}
where the extreme points are given by amplitude damping channels.

However, this  turns out  not to be  the case  because the $p$  in Eq.
(\ref{eq:naiv})  is  also  a   function  of  channel  parameters  that
determine the  extreme points.   Thus, varying the  `convex' parameter
$p$ shifts  the extreme  points.  The rather  complicated relationship
between  $p$ and  $\alpha,\nu,\mu$  is given  in Ref.   \cite{srisub}.
However, this functional dependence of the convex parameter on channel
properties implies the following result.
\begin{thm}
The set of all SGAD channels is not convex.
\label{thm:x}
\end{thm}
{\bf Proof.}  Assume that Eq.  (\ref{eq:naiv}) defines  a valid convex
set in the model for arbitrary  $p$ in the range $[0,1]$.  The channel
parameters  are temperature,  squeezing,  etc., which  may be  denoted
$x_i$ (for $1 \le i \le N$,  for any finite $N$).  Since $p = p(x_i)$,
each possible choice  of $p$ constitutes a constraint  to be satisfied
while keeping  fixed the  extreme points which  are also  functions of
$x_i$.  Clearly,  this is  impossible to satisfy  for any  finite $N$.
\hfill $\blacksquare$

\bigskip

It is an interesting question how the locus of the extreme points as
a function of  $p$ relates to the general  theory of area preserving
canonical transformations.

Since all unitary operations  are mapped to pure (maximally entangled)
Choi matrices, mixedness of the latter implies decoherence in the
channel.  This  suggests  that   the  degree  of  decoherence  can  be
quantified   by  the  amount   of  mixedness   of  the   Choi  matrix,
$C_{\cal{E}}$, Eq. (\ref{eq:choi}), Ref. \cite{geo}.
\begin{equation}
S = -\textrm{Tr}\left[ C_{\mathcal{E}} \log_2 C_{\mathcal{E}}\right].
\label{eq:chent}
\end{equation}
Likewise, since the noise acting on  one of the states will lead to
a reduction  in correlation between  the two states, the  degree of
separableness  of  the  Choi  matrix  also  can  be  considered  as  a
quantification  of  channel decoherence.   An  appropriate measure  of
entanglement is concurrence \cite{con}:
\begin{equation}
\mathcal{L}=\textrm{max}\left[0,\lambda_1-\lambda_2-\lambda_3-\lambda_4\right]
\label{eq:chcon}
\end{equation}
where $\lambda_1,\lambda_2,\lambda_3,\lambda_4$ are the eigenvalues of
$(\sigma_y \otimes \sigma_y)C_{\cal{E}} (\sigma_y \otimes \sigma_y)^T$
arranged  in  decreasing  order. 

\begin{figure}[h]
\begin{center}
\subfigure[~~Channel entropy (Eq. (\ref{eq:chent}))]{
\label{fig:chent}
\includegraphics[width=0.45\textwidth]{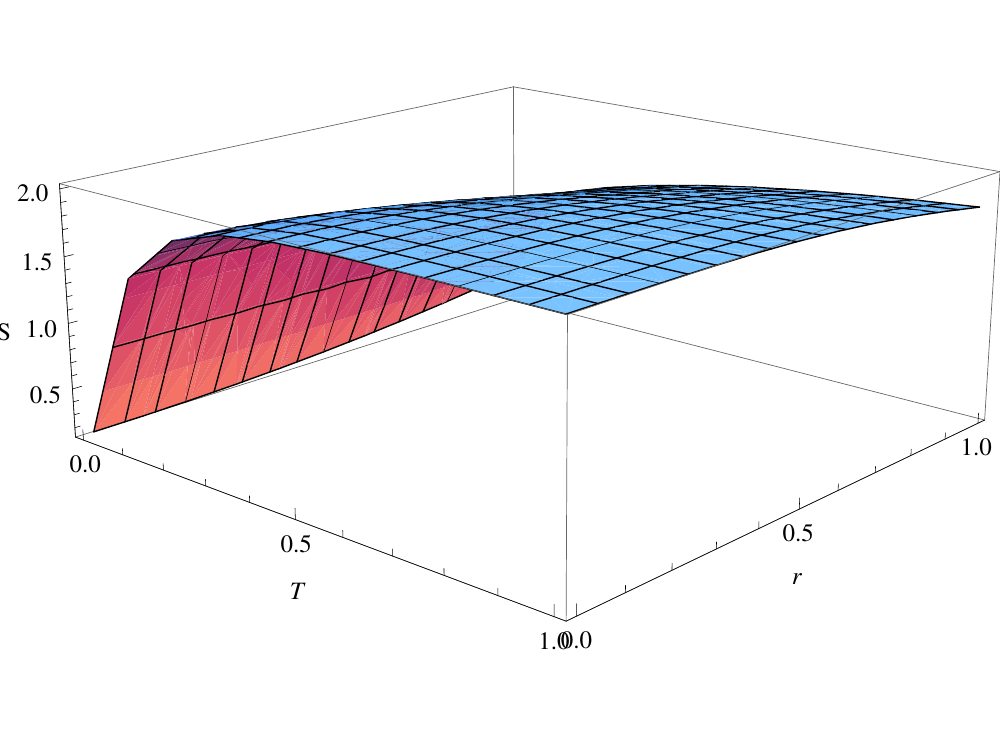}}
\subfigure[~~Channel concurrence (Eq. (\ref{eq:chcon}))]{
\label{fig:chcon}
\includegraphics[width=0.45\textwidth]{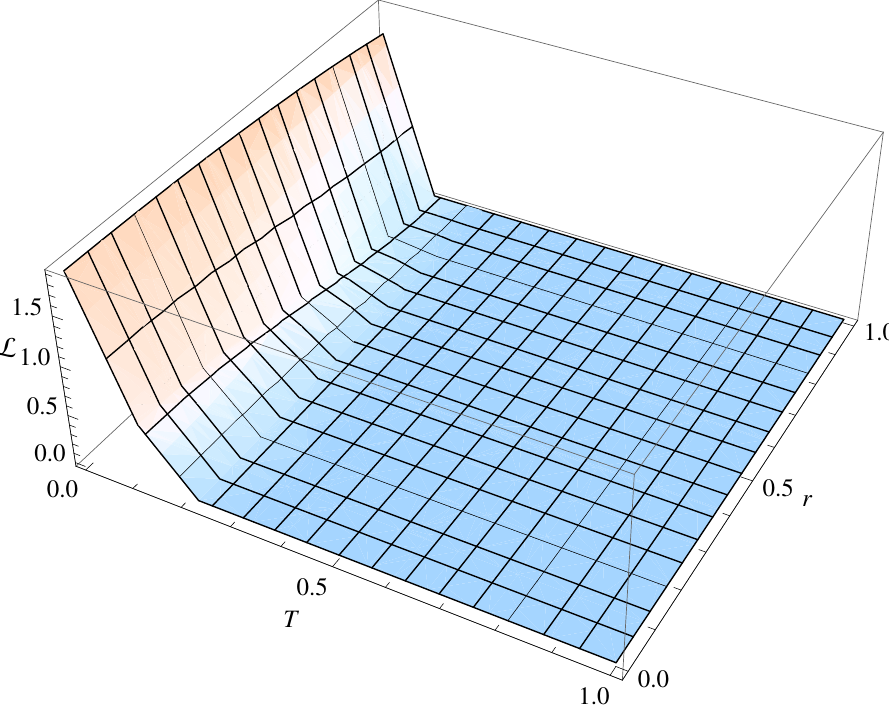}}
\end{center}
\caption{Two possible  quantifications of  the decoherence due  to the
  action  of a  SGAD channel,  as  function of  temperature ($T$)  and
  squeezing,   parametrized  by  $r$,   with  time   $t=0.5$,  frequency
  $\omega=0.01$,  bath parameter $\gamma_0 = 0.1$ and  squeezing angle
  $\Phi=0.3$ (in the units where $\hbar=k_B=1$). }
\end{figure}

As a quick  illustration,  for a  phase  flip channel  given by  Kraus
operator    elements   $\{\sqrt{p}I,   \sqrt{1-p}Z\}$    ($0\le   p\le
\frac{1}{2}$), it is  easily seen that the von  Neumann entropy of the
Choi  matrix  is  given  by  $H(p)$,  the  Shannon  entropy,  and  the
concurrence  by $1-2p$.   Thus  we find  that  as the  noise level  is
increased, so does the degree of mixedness and the separability of the
Choi  matrix. In Figures  (\ref{fig:chent}) and  (\ref{fig:chcon}), we
plot the von Neumann enropy and concurrence, respectively, of the Choi
matrix subjected  to a  SGAD channel.  As  expected we find  that they
indicate an  increase of decoherence  under increase of  $T$. However,
there are regions where  squeezing appears to suppress decoherence, as
seen from the Figure \ref{fig:chent}, near $T=1$.

\section{Gate and channel fidelities}

We here characterize  the performance of the SGAD  channel in terms of
parameters which, as noted earlier, should be unitarily invariant. One
such  quantity  is the  gate  fidelity  \cite{magesan}
\begin{equation}
g = \max_\rho F(\rho,{\cal E}(\rho)),
\end{equation}
the fidelity  of a  state with its  noisy version, maximized  over all
states,         where        fidelity         $F(\rho,\sigma)        =
\sqrt{\sqrt{\rho}\sigma\sqrt{\rho}}$.  Intuitively  it represents  how
well  a gate  performs  the  operation it  is  supposed to  implement.
Another is the average gate fidelity \cite{b,mn}:
\begin{equation}
g_{av} = \int_\rho F(\rho,{\cal E}(\rho)) \omega(\rho) d\rho,
\end{equation}
where  $\omega$ is  a suitable  uniform measure  over state  space.  A
closed   analytic   expression   exists   for  this,   due   to   Ref.
\cite{horodecki}, given by
\begin{equation}
g_{\rm av} = \frac{d + \sum_i |\textrm{Tr}(E_i)|^2}{d(d+1)}.
\label{eq:avgfid}
\end{equation}
Another similar parameter is teleportation distance \cite{vidal}.

A  related parameter,  introduced in  Ref.  \cite{srisub},  is channel
fidelity,  which  is  a measure  of  how  well  a gate  preserves  the
distinguishability of states:
\begin{equation}
\kappa \equiv \max_{\cal B}\chi({\cal B},{\cal E}),
\label{eq:chfid}
\end{equation}
where  $\chi({\cal B},{\cal  E})$  is  the Holevo  bound  for a  state
prepared in  basis elements  of basis ${\cal  B}$, subjected  to noise
${\cal   E}$.    Thus   $\kappa$  maximizes   the   distinguishability
(quantified  by the  Holevo bound)  over all  possible bases  (sets of
orthonormal states).   Clearly $\kappa \le  {\cal C}_1 \le  {\cal C}$,
where ${\cal C}_1$  is the product state channel  capacity, and ${\cal
  C}$, the  channel capacity maximized over  $n$-fold entanglement ($n
\rightarrow \infty$) \cite{cortese}.

$\kappa$  manifestly  possesses   unitary  invariance  because  it  is
computed from  the density operator directly, and  is thus independent
of the  tracing basis  used to obtain  the Kraus  operators.  Although
currently no  known closed expression exists for  channel fidelity, we
expect that its  behavior should be similar to  that of gate fidelity,
at least qualitatively. This  expectation is supported in a comparison
of the effect of temperate and squeezing on them, as discussed below.

\begin{figure}
\begin{center}
\subfigure[~~Average gate fidelity (Eq. (\ref{eq:avgfid}))]{
\label{fig:avgfid}
\includegraphics[width=0.45\textwidth]{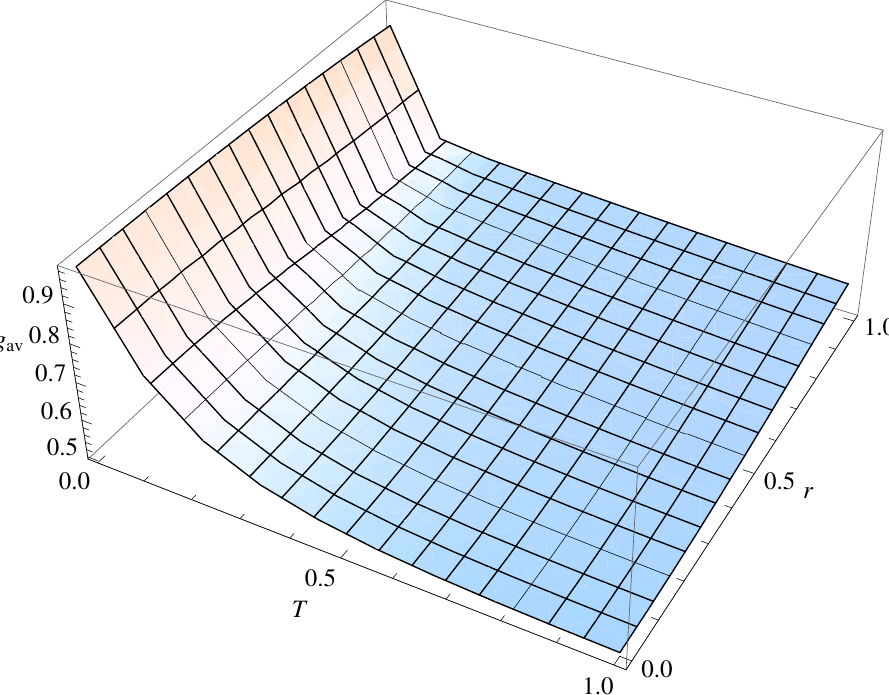}}
\subfigure[
 ~~Channel fidelity (Eq. (\ref{eq:chfid}))]{
\label{fig:chfid}
\includegraphics[width=0.45\textwidth]{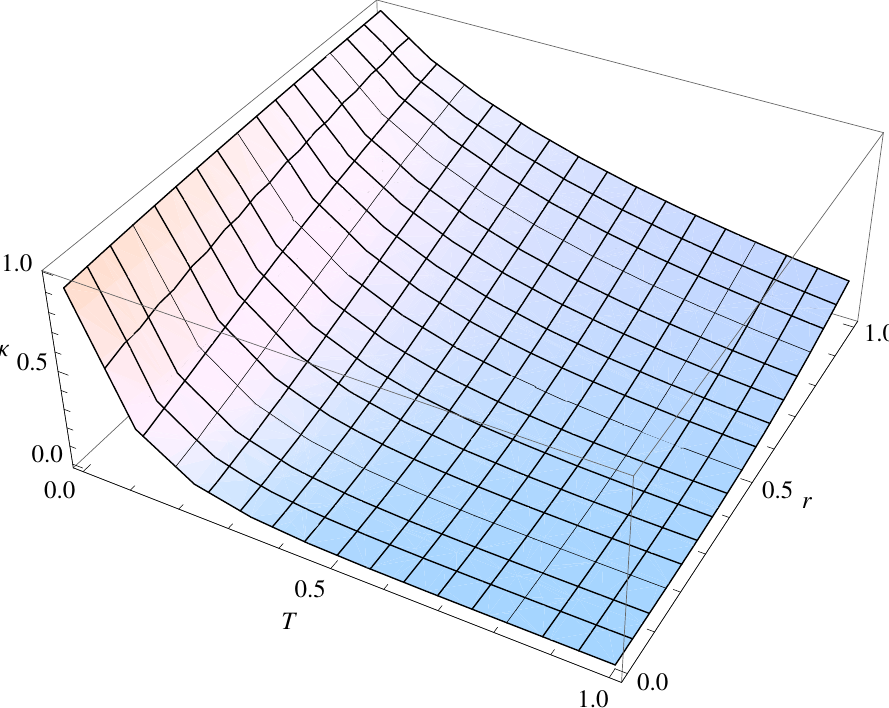}}
\end{center}
\caption{SGAD channel properties, as function of temperature ($T$) and
  squeezing,   parametrized  by  $r$,   with  time   $t=0.5$,  frequency
  $\omega=0.01$, bath parameter  $\gamma_0=0.1$ and  squeezing angle
  $\Phi=0.3$ for (a) and  $\Phi=0.0$ for (b) (in the units  where $\hbar=k_B=1$). Both quantities show
  a counter-intuitive rise with squeezing.}
\end{figure}

A  plot of  $g_{\rm  av}$ for  the  SGAD channel  is  given in  Figure
(\ref{fig:avgfid}).   While  at  low enough temperature,  squeezing  reduces
average gate fidelity $g_{\rm av}$,  a range of temperature is seen to
exist,  in which  squeezing causes  an  increase in  the quantity.   A
similar  counter-intuitive  reduction of  $\kappa$  with squeezing  is
depicted in Figure \ref{fig:chfid}.

What is remarkable  is that even in regimes  where noise increases, as
indicated    by   Figures    \ref{fig:chent},    \ref{fig:chcon}   and
\ref{fig:avgfid}),   obtained  by   increasing   squeezing  at   fixed
temperature, the distinguishability  of intially orthogonal states, as
given by  channel fidelity  in Figure \ref{fig:chfid}),  increases. We
confirmed  this  behavior  by  employing the  trace  distance  measure
instead of  channel fidelity, obtaining the same  pattern.  It follows
that in  this signaling ensemble,  inspite of the diffusion  caused by
noise,  the intially  orthogonal  states continue  to  enjoy a  nearly
orthogonal  support. Invariably,  this feature  happens only  when the
increase in decoherence is due to increase in squeezing $r$ at a given
temperature.

\section{Discussions and Conclusions}

Single-qubit channels have been studied under the two broad classes of
AD and generalized depolarizing  channels, which are fairly exhaustive
in  real life  situations.  Two  of  the authors  had earlier  derived
\cite{srisub} an  operator sum representation of the  SGAD channel, by
generalizing the  GAD channel.  A different  derivation, that exploits
the Choi channel-state isomorphism  was presented here, along with the
unitary operation relating it to the previous derivation.

There is  a rich structure  to be explored  by the isomorphism.   As a
small part of larger work that may be undertaken here, we characterize
the difference in the geometry and rank of these channel classes.  The
degree of decoherence of the  qubit channel is quantified according to
the amount of mixedness, as  quantified by the von Neumann entropy, or
separability, quantified  by the absence  of concurrence, of  the Choi
matrix.  

Whereas  the  generalized   depolarizing  channels  possess  a  convex
structure and  form a 3-simplex, the  AD class channels  lack a convex
set  as  seen  from  an  ab  initio perspective,  and  are  thus  more
complicated  to  study.   Further,   where  the  rank  of  generalized
depolarizing  channels can  be any  positive integer  upto 4,  that of
amplitude damping ones is either  2 or 4.  Various channel performance
parameters  can be  used  to  bring out  the  different influences  of
temperature and  squeezing in dissipative channels.   In particular, a
noise range  in terms  of $r$ and  $T$ was identified  where initially
orthogonal  states  prepared  in  a  suitable basis  can  become  more
distinguishable  inspite   of  decohering.  This   happens  only  when
squeezing is increased, rather than temperature.

\end{document}